\documentclass[preprint,12pt]{elsarticle}

\usepackage{graphicx}
\usepackage{color}
\usepackage[colorlinks,bookmarks=false,citecolor=blue,linkcolor=red,urlcolor=blue]{hyperref}

\usepackage{amssymb,amsbsy,amsmath}

\newcommand {\mofe} {Mo$_{72}$Fe$_{30}$}
\newcommand {\mov} {Mo$_{72}$V$_{30}$}
\newcommand {\mocr} {Mo$_{72}$Cr$_{30}$}

\newcommand {\wfe} {W$_{72}$Fe$_{30}$}
\newcommand {\wv} {W$_{72}$V$_{30}$}

\newcommand{\op}[1]{%
    \fontdimen12\textfont3=2pt\fontdimen12\scriptfont3=1.4pt%
    \!\null\mathop{\vphantom{#1}\smash{#1}}\limits_{\sim}\null\!}

\journal{J. Magn. Magn. Mater.}

\begin{document}

\begin{frontmatter}



\title{Large-scale numerical investigations of the antiferromagnetic Heisenberg icosidodecahedron}


\author{J{\"o}rg Ummethum\fnref{bi}}
\author{J{\"u}rgen Schnack\corref{cor1}\fnref{bi}}
\ead{jschnack@uni-bielefeld.de}
\author{Andreas M. L{\"a}uchli\fnref{ins}}
\cortext[cor1]{corresponding author}
\address[bi]{Dept. of Physics, Bielefeld University, P.O. box
  100131, D-33501 Bielefeld, Germany}
\address[ins]{Inst. f. Theoretische Physik, Innsbruck
  University, Technikerstr. 25, 6020 Innsbruck, Austria}

\begin{abstract}
We present up to date investigations of the 
antiferromagnetic Heisenberg icosidodecahedron by means of the
Density Matrix Renormalization Group method. We compare our
results with modern Correlator Product State as
well as Lanczos calculations.
\end{abstract}


\end{frontmatter}


\section{Introduction}
\label{sec-1}

\begin{figure}[ht!]
	\centering
	\begin{minipage}[c]{0.5\textwidth}
		\includegraphics[width=1.0\textwidth]{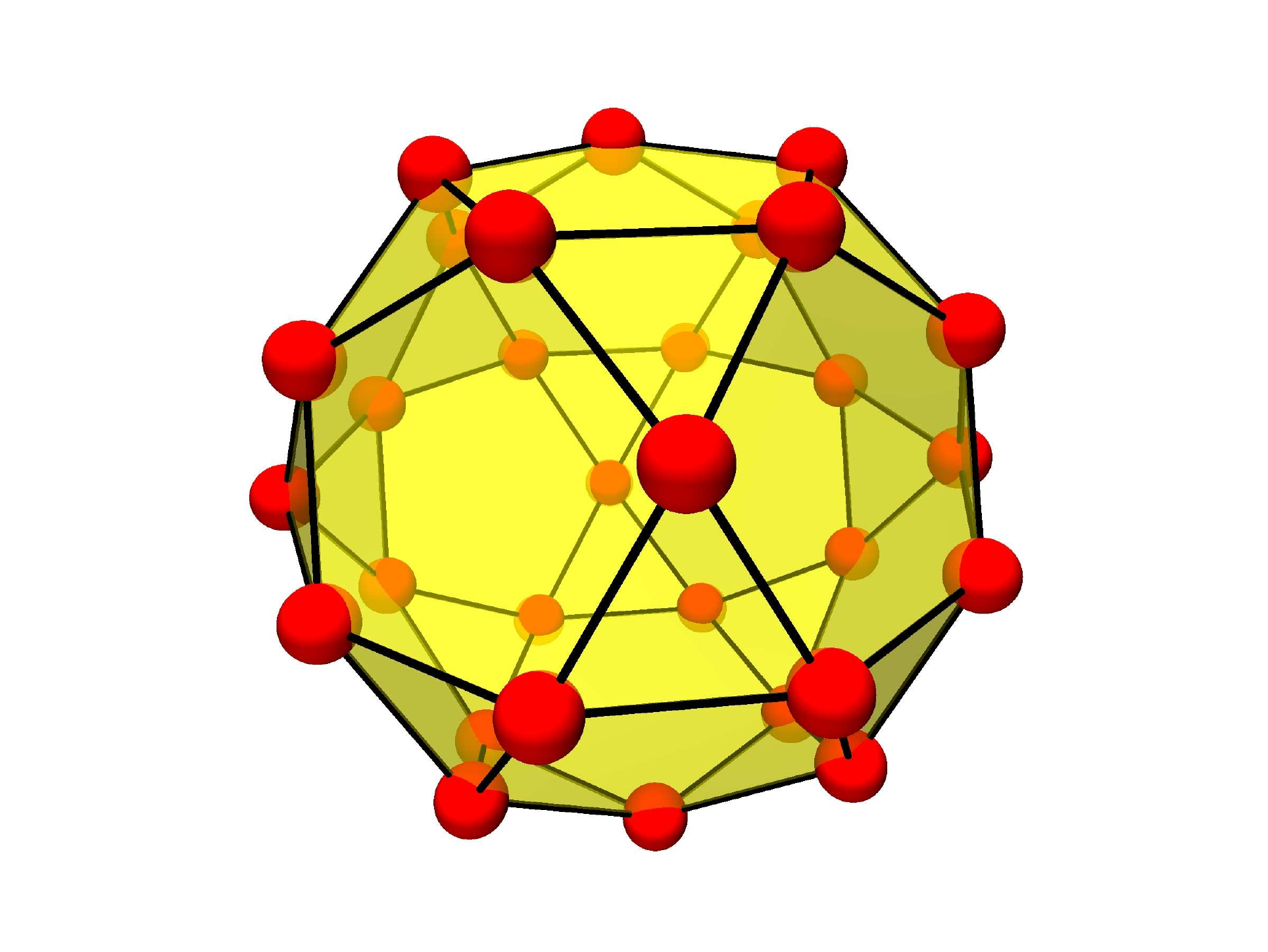}
	\end{minipage}
	\begin{minipage}[c]{0.45\textwidth}	
		\includegraphics[width=0.8\textwidth]{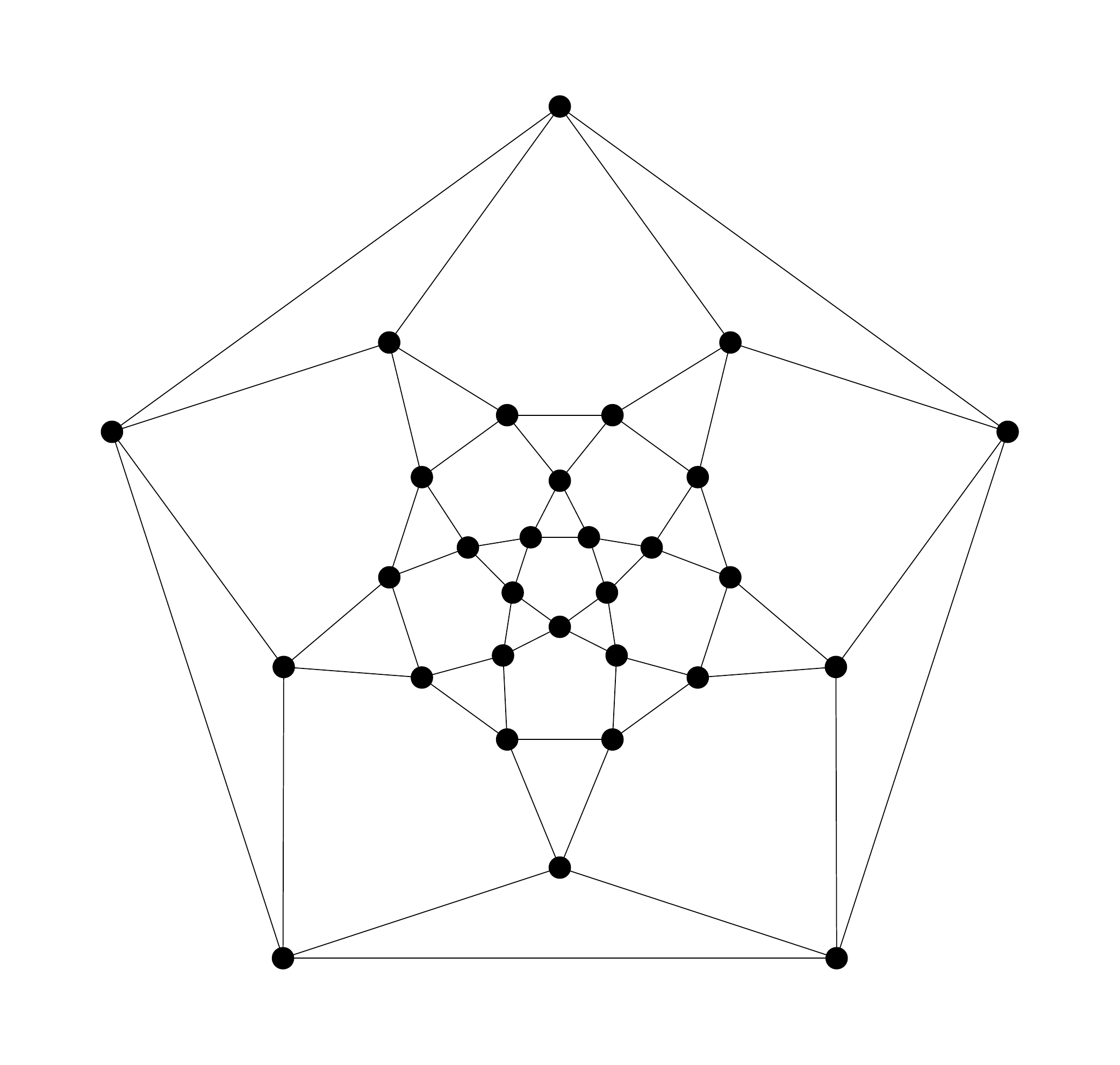}
	\end{minipage}	
	\caption{Structure of the icosidodecahedron: the black
	bullets correspond to the spin positions and the lines
	to interaction paths between them. The right part of the
	figure shows the two-dimensional projection.} 
	\label{fig:ico_joerg-2}
\end{figure}

Thanks to advanced chemical strategies there exist several
chemical realizations of icosidodecahedral 
magnetic molecules: \mofe 
\cite{Fe30-2001}, \wfe \cite{Tetal:AngewChemIntEd2010} (both
$s=5/2$), \mocr \cite{Cr30-2007} ($s=3/2$), \mov
\cite{V30-2005,BKH:ChemCommun2005}, and \wv
\cite{Tetal:ChemCommun2009} (both $s=1/2$). These molecules are
some of the largest magnetic molecules synthesized to date
\cite{GSV:Book2011}. Icosidodecahedral magnetic
molecules are of special interest because they are highly
symmetric, frustrated, exist with different spin quantum
numbers, and are a kind of finite-size version of the Kagom\'e
lattice antiferromagnet \cite{RLM:PRB2008}. Fig.~\ref{fig:ico_joerg-2} shows the
structure of the icosidodecahedron. It is an Archimedian Solid with 12
pentagons and 20 triangles, which means
that it is geometrically frustrated \cite{Sch:DaltonTrans2010}. 

Experimental investigations of these molecules are in most cases
measurements of the susceptibility as a function of temperature
\cite{Fe30-2001,Tetal:AngewChemIntEd2010,Cr30-2007,V30-2005,BKH:ChemCommun2005,Tetal:ChemCommun2009}
or magnetic field \cite{Setal:PRB2008,KMvSSDD:CoordChemRev2009},
or magnetization as a function of the applied magnetic field
\cite{Fe30-2001,KMvSSDD:CoordChemRev2009,SLM:EPL2001}. These
experimental investigations show that the icosidodecahedral
magnetic molecules are antiferromagnetic with a nonmagnetic
ground state. Other experimental techniques that were applied to
these molecules are NMR and $\mu$SR
\cite{Jetal:JApplPhys2002,Metal:JMagnMagnMater2004,Letal:PRB2007},
INS \cite{Getal:PRB2006}, diffuse (elastic) neutron scattering,
as well as specific heat measurements \cite{FKRSMB:NewJPhys2010}.

These molecules are usually modeled using a simple Heisenberg
model
\cite{Fe30-2001,Tetal:AngewChemIntEd2010,Cr30-2007,V30-2005,Tetal:ChemCommun2009,Getal:PRB2006,FKRSMB:NewJPhys2010,SNSHLK:PRL2005,AxL:PRB2001,Wal:PRB2007,ScW:EurPhysJB2010,HaS:JPhysSocJpn2004,KTF:JPhysConfSer2009}.
Anisotropic terms were considered in
Refs.~\cite{HaS:JPhysSocJpn2004,CeZ:ProgTheorPhys2005}. Bond
disorder and distortions (i.e., more than just one exchange
constant in the Heisenberg Hamiltonian) were investigated in
Refs.~\cite{Setal:PRB2008,KTF:JPhysConfSer2009}. However, since
the icosidodecahedral molecules comprise $N=30$ spins, the
numerical exact calculation of $T=0$ properties is possible only
for the $s=1/2$ case
\cite{RLM:PRB2008,SRS:JMagnMagnMater2005,SSRS:EurPhysJB2001}.
Quantum Monte Carlo suffers from the negative-sign problem so
that small temperatures are not feasible
\cite{Cr30-2007,V30-2005,Tetal:ChemCommun2009}. Thermodynamical
properties at $T>0$ can for $s=1/2$ also be calculated
quasi-exactly using the finite-temperature Lanczos method
\cite{ScW:EurPhysJB2010,KTF:JPhysConfSer2009,PhysRevB.49.5065}. For
$s>1/2$, approximations are needed. For $s=3/2$ and
$s=5/2$ systems, the classical Heisenberg model was used
together with efficient classical Monte Carlo algorithms
\cite{Fe30-2001,Tetal:AngewChemIntEd2010,Setal:PRB2008,SNSHLK:PRL2005,AxL:PRB2001,HaS:JPhysSocJpn2004,ToS:PRB2007}.
However,
such an approximation is inappropriate at very low
temperatures. The rotational band approximation was used in
Refs.~\cite{Fe30-2001,SLM:EPL2001,Getal:PRB2006,FKRSMB:NewJPhys2010,Wal:PRB2007,ScL:PRB2000,SLS:CMP09}.
Although being a quantum mechanical approximation it misses
important features of
frustrated systems such as magnetization jumps and plateaus or
low-lying singlets in the spectrum
\cite{SRS:JMagnMagnMater2005}. Another approximation applied to
the icosidodecahedron is spin-wave theory
\cite{Wal:PRB2007,CeZ:ProgTheorPhys2005}. However, as for the
rotational band and the classical approximation, it is not clear
how accurate this approximation is. 

It has to be emphasized here that although there exist many
theoretical studies on the icosidodecahedron for $s>1/2$,
accurate numerical calculation for the \textit{full} Heisenberg
model are very rare. The Density Matrix Renormalization Group
(DMRG) method allows for treating the full Heisenberg
Hamiltonian but in a reduced Hilbert space \cite{Whi:PRL1992,Sch:RMP05}. It 
relies on a controlled truncation of the Hilbert space and
allows for the estimation of the accuracy so that it seems to be
suited for these systems. In Ref.~\cite{ExS:PRB2003} the DMRG
method has already been applied to the Heisenberg
icosidodecahedron with $s=5/2$. However, only up to $m=120$
density matrix eigenstates were used, so that the accuracy of
the results is rather limited for such a complicated system with
a geometry that is not favorable for the DMRG method. 

In this article we apply the DMRG method to the antiferromagnetic
Heisenberg icosidodecahedron. We focus on the calculation of the
lowest energies in subspaces of total magnetic quantum number
$M$ which allow for a calculation of the $T=0$ magnetization
curve and also gaps which might be of importance for
spectroscopic methods such as, e.g., 
Inelastic Neutron Scattering (INS). These results are compared with
very recent variational Monte Carlo calculations using
correlator product states (CPS) by Neuscamman and Chan in
Ref.~\cite{NeC:arXiv2012}. Finally, for the case $s=1/2$ we also
calculate the dynamical correlation function $S_\mathrm{loc}^{z}(\omega)$
using the Dynamical DMRG (DDMRG) \cite{Jec:PRB02}.

\section{DMRG results}
\label{sec-2}

The DMRG technique is best suited for open one-dimensional chain
systems but can be applied to systems with an arbitrary
geometry. The icosidodecahedron can be viewed as a
two-dimensional lattice on a sphere (similar to the Kagome
lattice, see \cite{RLM:PRB2008}), i.e., with periodic boundary
conditions. This means that the convergence is much slower than
in one-dimensional systems \cite{ExS:PRB2003}. But since DMRG is a variational
method, it is clear that the ground state -- or the lowest
energy in a subspace -- is the better the lower the
corresponding energy is. Also, the truncated weight $\Delta w$
offers the possibility to judge the quality of the results and
an extrapolation to zero truncated weight (or
$m\rightarrow\infty$) might be possible. 

For the investigations throughout this article we employ the
Heisenberg Hamiltonian
\begin{eqnarray}
\label{E-3-1}
\op{H}
&=&
J\;
\sum_{i<j}\;
\op{\vec{s}}_i \cdot \op{\vec{s}}_j
+
g\, \mu_B\, B\,
\sum_{i}\;
\op{s}^z_i
\end{eqnarray}
with antiferromagnetic isotropic nearest neighbor exchange $J$
only.

\subsection{Numbering of the spins}

When DMRG is applied to spin systems that are not
one-dimensional, the usual way is to map the system on a
one-dimensional chain with long-range interactions, i.e., to
number the spins of the lattice \cite{StW:AnnRev2012}. However,
if not very simple systems such as, e.g., ladders are
investigated, it is not clear, which numbering is best
suited. Such a problem also occurs, when DMRG is applied in the
context of quantum chemistry, where models similar to the
Hubbard model with long-range interactions appear and the
ordering, i.e., the numbering of the orbitals is relevant
\cite{ChH:JChemPhys2002,MLPF:JChemPhys2003,LeS:PRB2003,MHR:JChemPhys2005,RNW:ChemPhys2006}.
Since long-range interactions diminish the accuracy of DMRG
(cf. Ref.~\cite{LiP:PRB1994}) it is clear that a good
ordering needs to minimize such long-range interactions. 

\begin{figure}[ht!]
	\begin{minipage}[c]{0.5\textwidth}
		\includegraphics[width=1.0\textwidth]{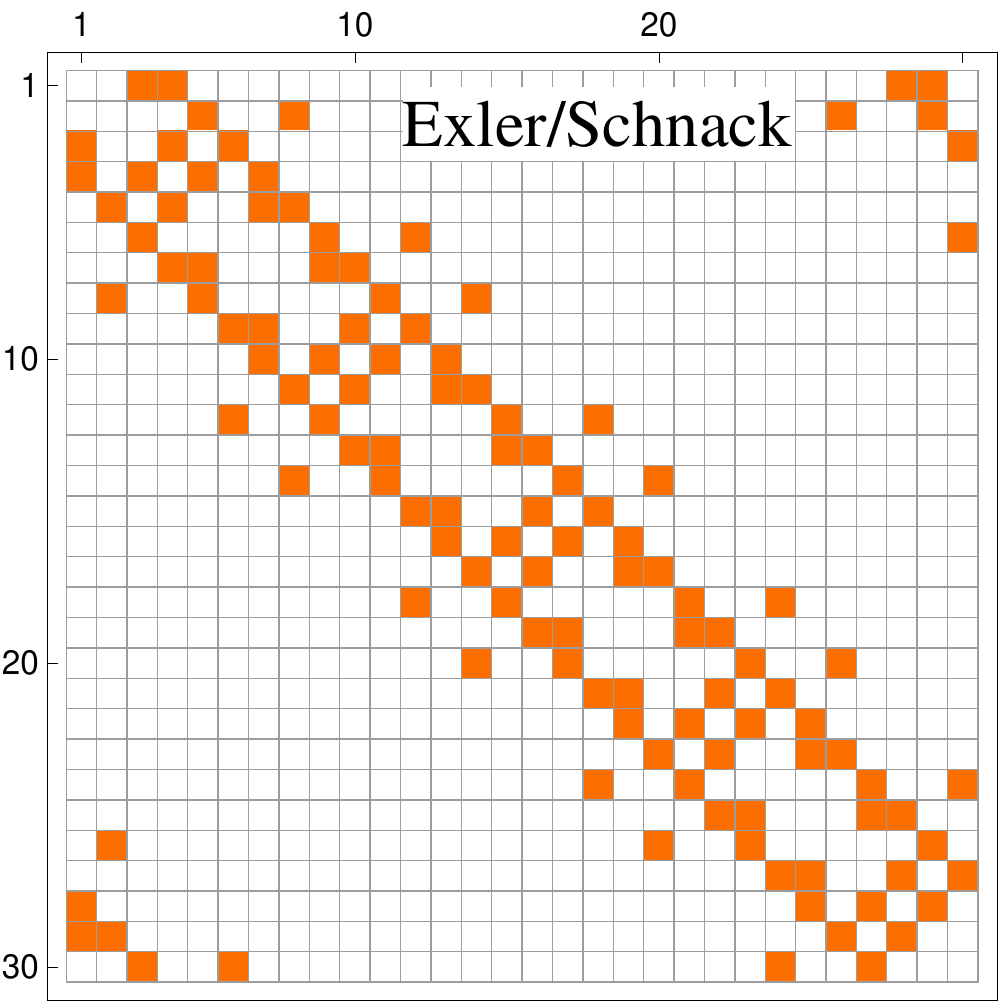}
	\end{minipage}
	\begin{minipage}[c]{0.5\textwidth}
		\includegraphics[width=1.0\textwidth]{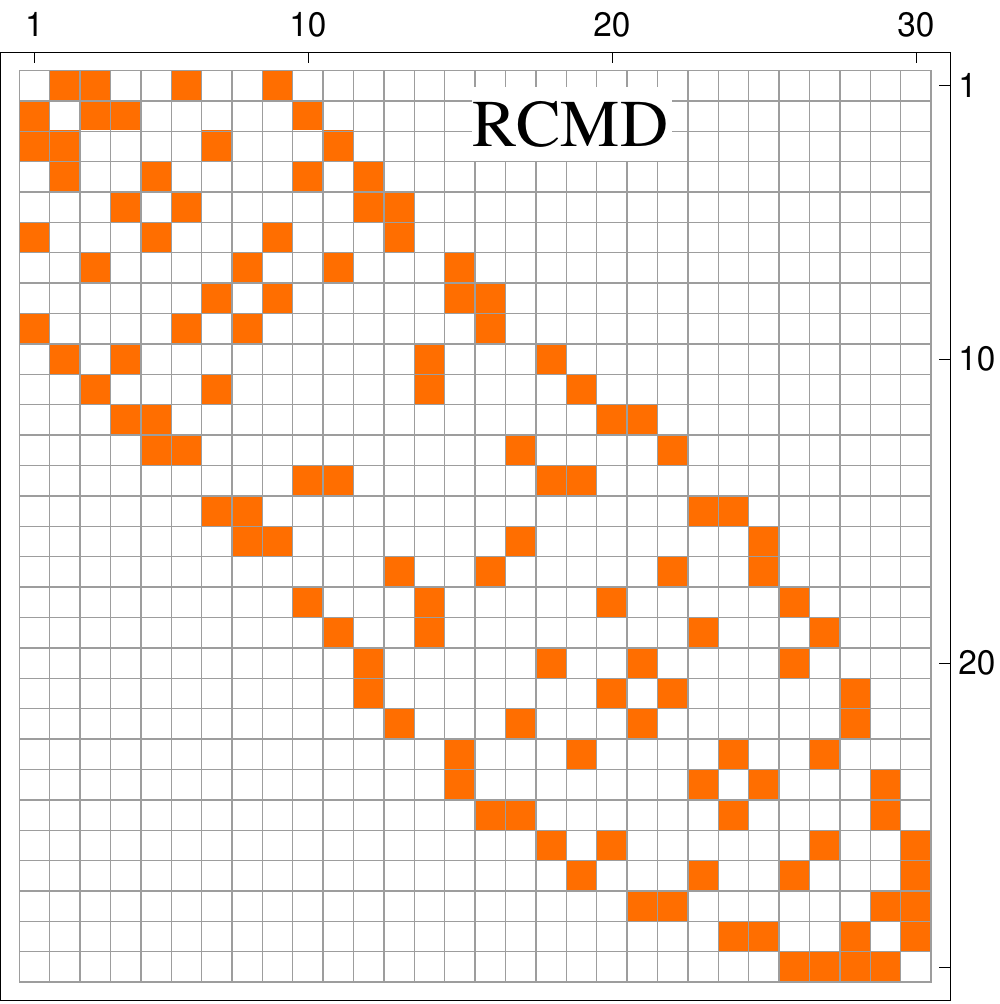}
	\end{minipage}
	\begin{minipage}[c]{0.5\textwidth}
		\includegraphics[width=1.0\textwidth]{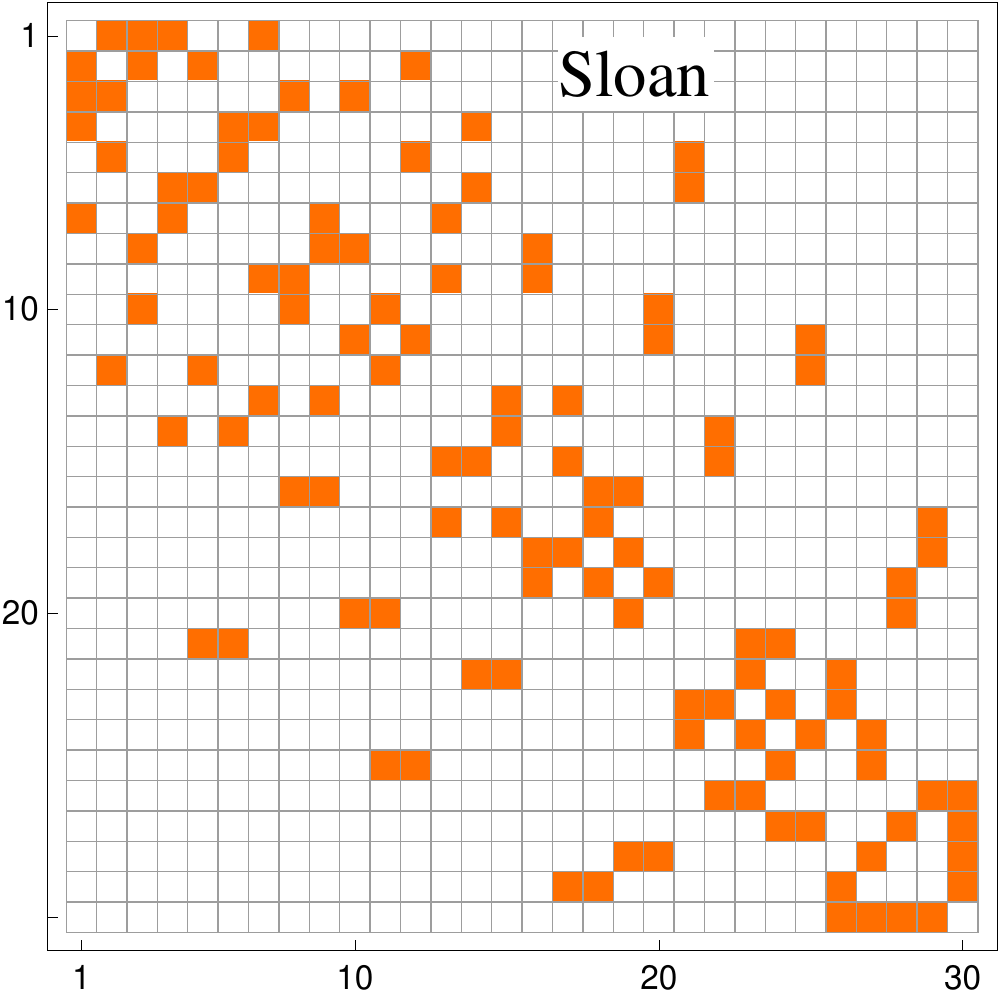}
	\end{minipage}
	\begin{minipage}[c]{0.5\textwidth}
		\includegraphics[width=1.0\textwidth]{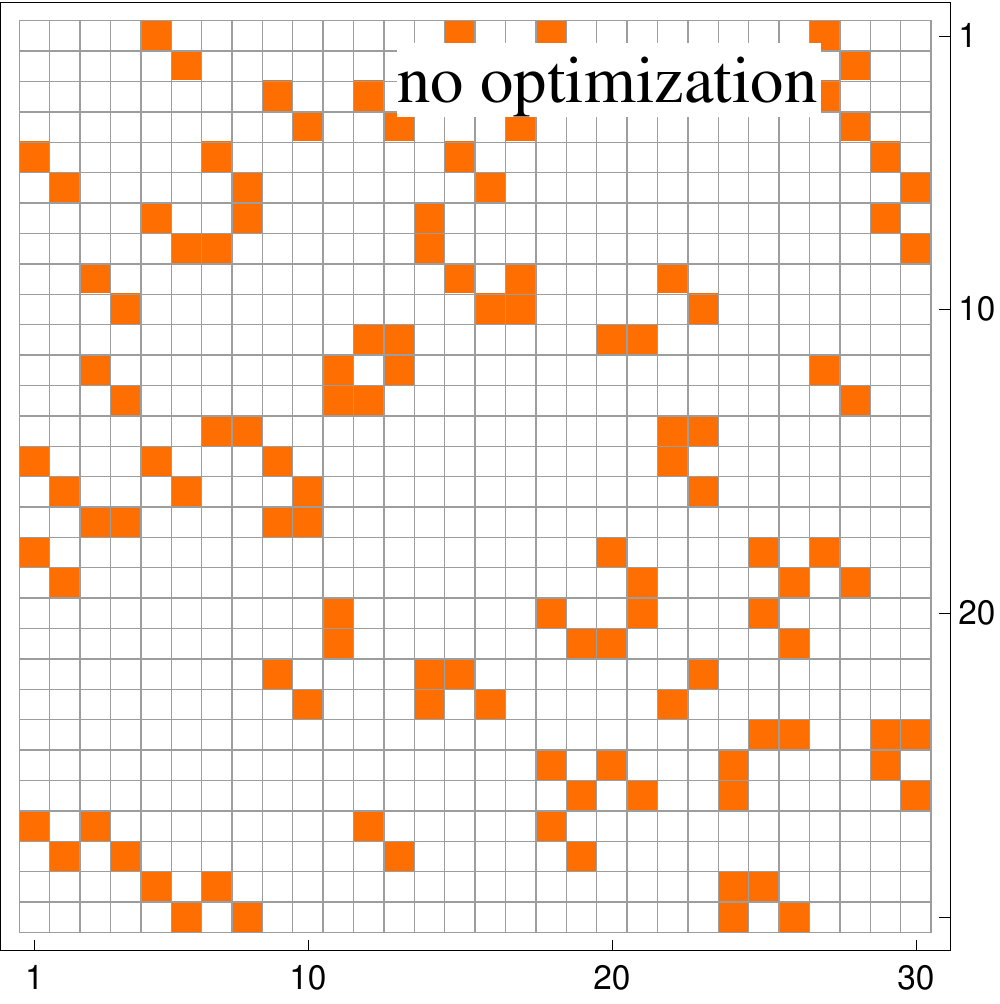}
	\end{minipage}
	\caption{Depiction of the coupling matrix $(J_{ij})$ for
	four different numberings of the vertices of the
	icosidodecahedron. Nonzero entries are denoted by the
	orange squares. Top left: numbering according to Exler
	and Schnack (see Ref.~\cite{ExS:PRB2003}); top right:
	result of the RCMD algorithm; bottom left: result of the
	Sloan algorithm; bottom right: unoptimized numbering.} 
	\label{fig:ico-numbering}
\end{figure}

We have tested several numberings for the icosidodecahedron. The
resulting coupling matrices $J_{ij}$ are shown in
Fig.~\ref{fig:ico-numbering}. The numbering used by Exler and
Schnack in an earlier investigation \cite{ExS:PRB2003}
(see top left of Fig.~\ref{fig:ico-numbering}) gives a very
regular ``interaction pattern'' with rather-short-ranged
interactions, but the ``periodic boundaries'', i.e.,
interactions between the first and the last spins, are clearly
not optimal for the DMRG algorithm with two center sites. As
proposed in Ref.~\cite{ChH:JChemPhys2002}, we have used a
variant of the reverse Cuthill-McKee algorithm
\cite{RCM-1,RCM-2}, the RCMD algorithm, which aims to number the
vertices such that the bandwidth of the matrix is minimized. We
have also used the Sloan algorithm \cite{Sloan} which minimizes
the ``envelope size'', i.e., the sum of the ``row
bandwidths''. (The bandwidth is the maximum of the row
bandwidths.) We have used these algorithms as implemented in
Mathematica \cite{MinimumBandwidthOrdering}. The figure also
shows an unoptimized numbering for comparison. 

\begin{figure}[ht!]
	\centering
		\includegraphics[width=0.65\textwidth]{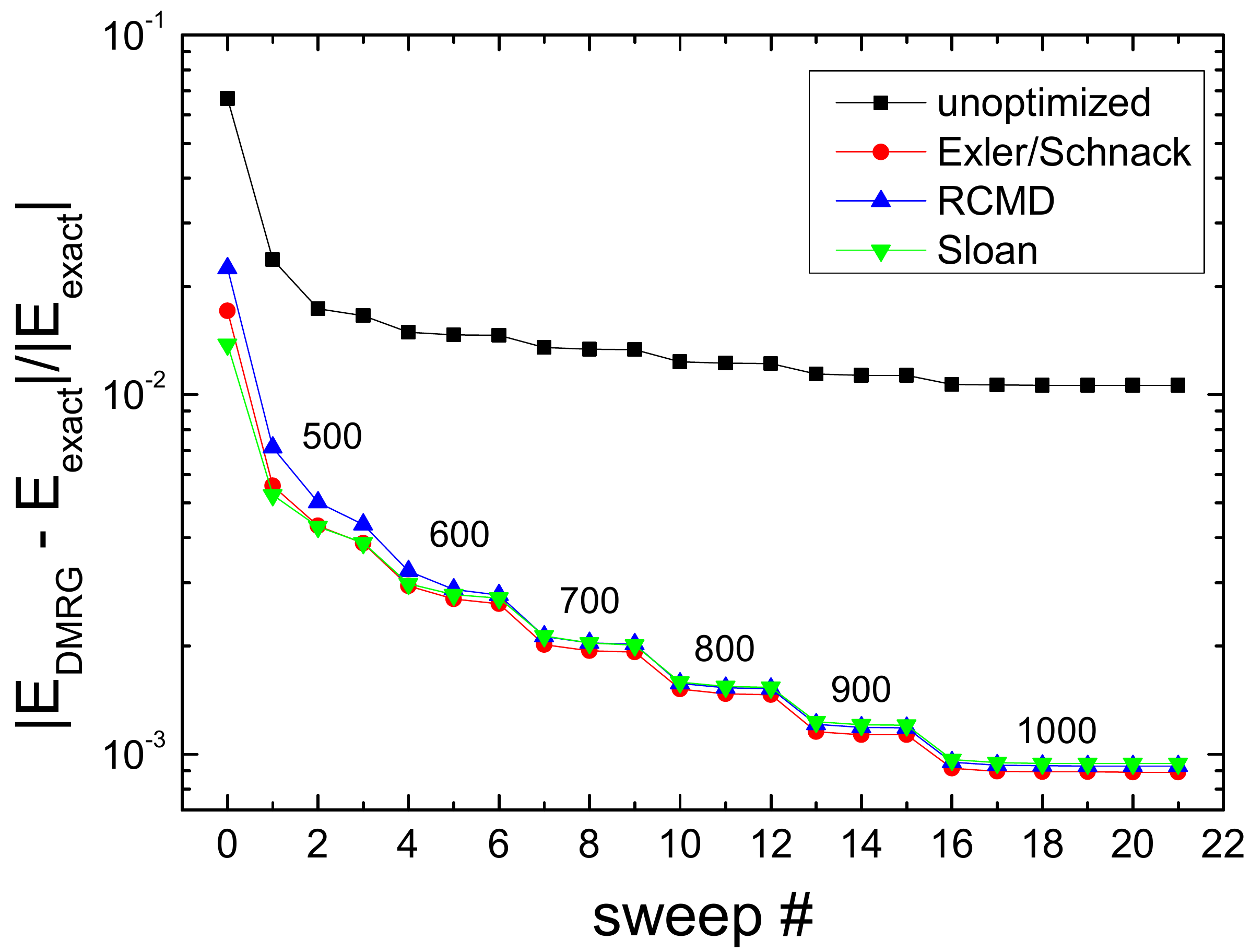}
	\caption{DMRG results for different numberings (see
	Fig.~\ref{fig:ico-numbering}) of the spins sitting on
	the vertices of the icosidodecahedron. The plot shows
	the error in the ground state energy as obtained by DMRG
	and as a function of the DMRG sweep. The numbers above
	the symbols denote the number of kept density matrix
	eigenstates for the sweep.} 
	\label{fig:ico-ordering-results}
\end{figure}

The results of DMRG calculations (using the ALPS DMRG code
\cite{ALPS2007}) for the different spin numberings are shown in
Fig.~\ref{fig:ico-ordering-results}. We have calculated the
ground state energy of the $s=1/2$ icosidodecahedron with an
increasing number of kept density matrix eigenstates ($m$) so
that the convergence can be investigated and a comparison with
the exact ground state energy (see Ref.~\cite{RLM:PRB2008}) is
possible. One can see that the different optimized numberings
(Exler/Schnack, RCMD, and Sloan) give almost identical results
whereas the unoptimized numbering gives much worse
results. These results show that 
a ``good'' numbering of the spins is absolutely essential if the
DMRG method is applied to a spin system with a complicated
structure. For the following results we have always used the
numbering as proposed by Exler and Schnack.

\subsection{Lowest energy eigenvalues and magnetization curves}

As a next step we have calculated the lowest energies in the $M$
subspaces for the icosidodecahedron with $s>1/2$ using DMRG. The
results for the  $s=1/2$ system already showed that DMRG is able
to produce very accurate results for this system with relative
errors smaller than $10^{-3}$.

\begin{figure}[ht!]
	\begin{minipage}[c]{0.5\textwidth}
		\includegraphics[width=1.0\textwidth]{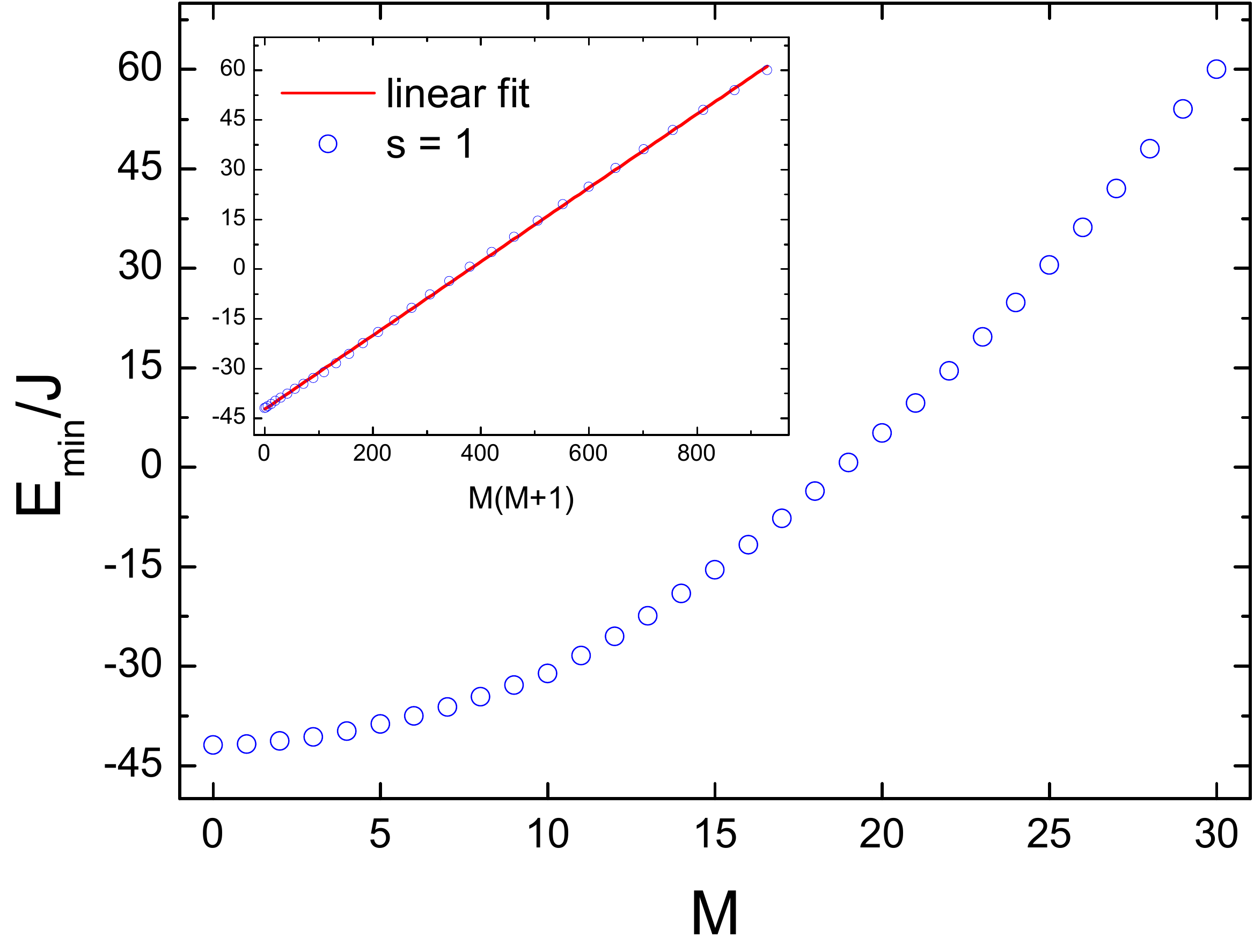}
	\end{minipage}
	\begin{minipage}[c]{0.5\textwidth}
		\includegraphics[width=1.0\textwidth]{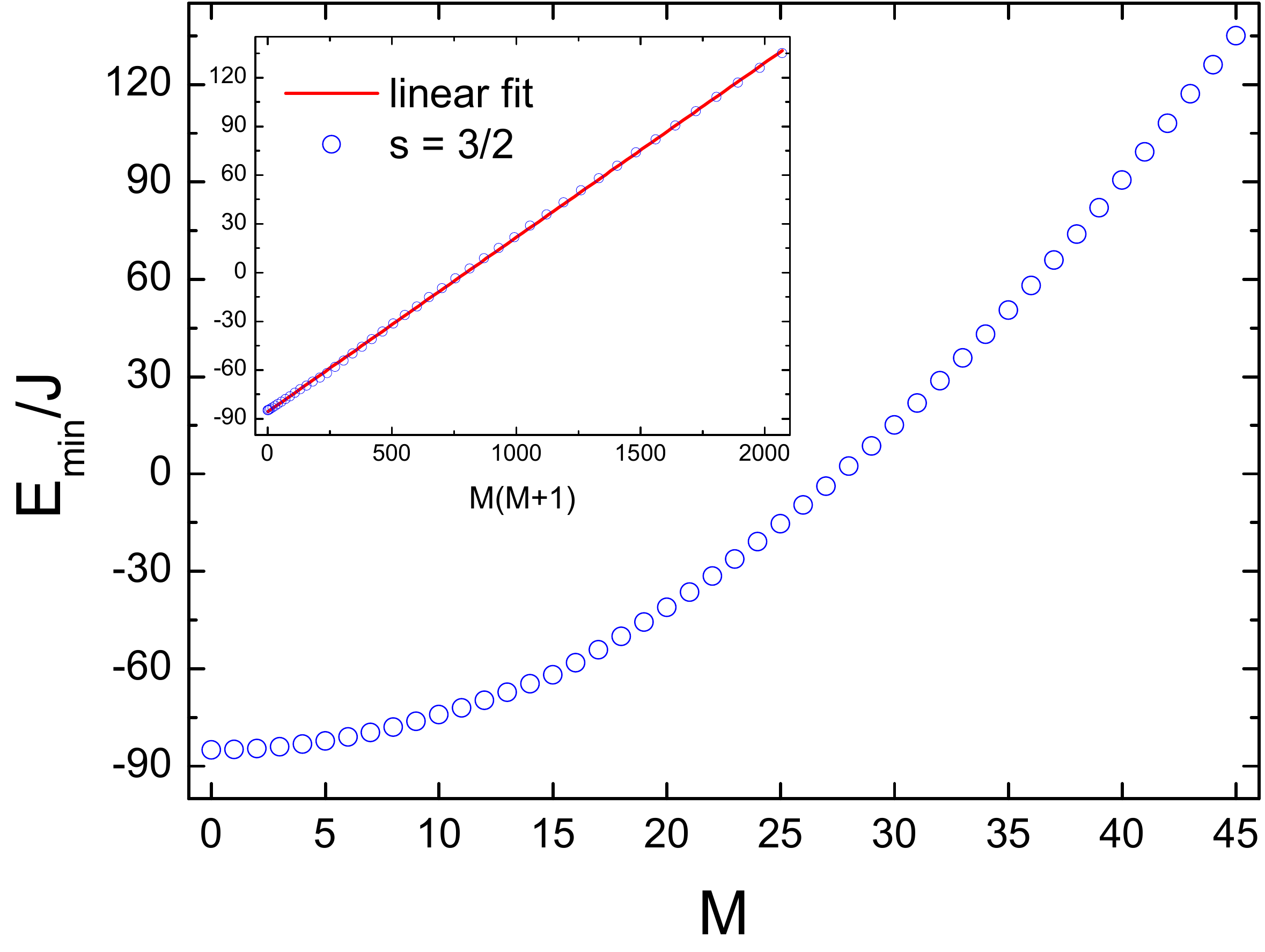}
	\end{minipage}
	\caption{Lowest energy eigenvalues in the subspaces of
	total magnetic quantum number $M$ as obtained by DMRG
	calculations. The eigenvalues for the smallest
	subspaces, i.e., for large $M$, were calculated using
	the Lanczos algorithm (calculations performed by
	J. Schnack, private communication). For the DMRG
	calculations, the ALPS DMRG code was employed
	\cite{ALPS2007}. For the $s=1$ system we have kept
	$m=2500$ density matrix eigenstates in all DMRG
	calculations. For the $s=3/2$ system we have kept
	$m=2000$ states for the subspaces up to $M=4$, $m=1500$
	states for the subspaces $5\leq M\leq 23$, and 1000
	states for the subspaces $M>23$.} 
	\label{fig:ico-rotbands}
\end{figure}

Fig.~\ref{fig:ico-rotbands} shows the lowest energy eigenvalues
in the subspaces of total magnetic quantum number $M$ for the
icosidodecahedron with $s=1$ and $s=3/2$ as obtained by DMRG and
-- for the large-$M$ subspaces ($M>18$ for $s=1$ and $M>33$ for
$s=3/2$) -- Lanczos calculations. We have used up to $m=2500$ density
matrix eigenstates for the $s=1$ case and up to $m=2000$ for the
$s=3/2$ case. The largest truncated weight within a
sweep is of the order of $7\cdot 10^{-4}$ for the $M=0$ subspace
of the $s=1$ icosidodecahedron and of the order of $4\cdot
10^{-4}$ for the $s=3/2$ case. That the truncated weight for the
$s=1$ icosidodecahedron is larger than for $s=3/2$ although more
states have been used for $s=1$ indicates that it cannot be
reliably used for a quantitative estimate of the error. The
reason for this behavior might be that the results are not yet fully
converged for the value of $m$ that we have used, although we have
carried out up to $60$ sweeps for the calculations.

\begin{figure}[ht!]
	\centering
		\includegraphics[width=0.65\textwidth]{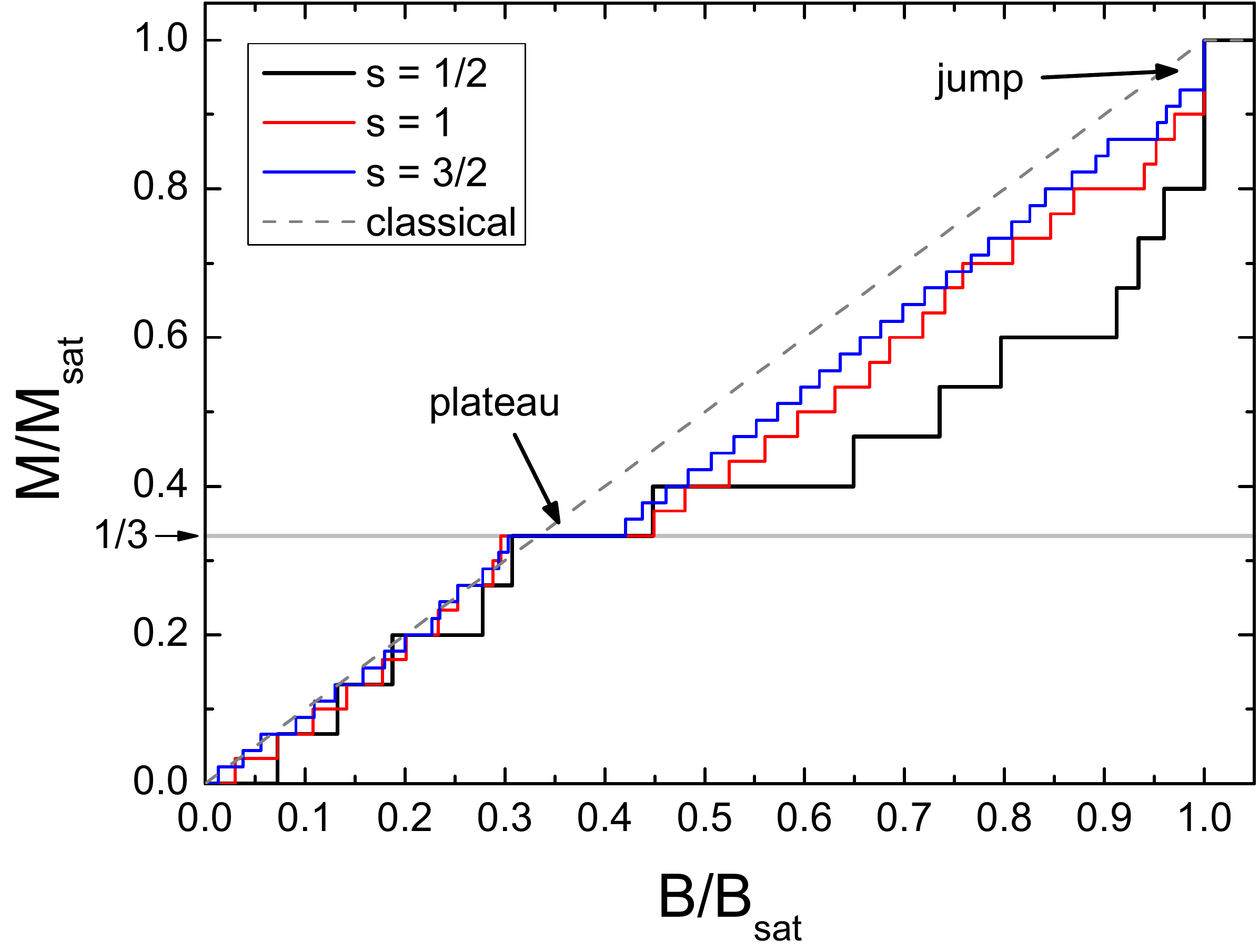}
	\caption{Zero-temperature magnetization curves of the
	icosidodecahedron as obtained by DMRG ($s=1$, $s=3/2$)
	calculations (cf. Fig.~\ref{fig:ico-rotbands}). The data
	for $s=1/2$ is taken from Ref.~\cite{RLM:PRB2008}. The
	plot also shows the classical result
	\cite{AxL:PRB2001}. The data is normalized to the
	saturation field and magnetization.} 
	\label{fig:ico-magnetization-curves-1}
\end{figure}

The rotational band model predicts a behavior of the form
$E_{\rm min}(M)=aM(M+1)+b$, i.e., a parabolic dependence
\cite{SLM:EPL2001}. The 
insets of Fig.~\ref{fig:ico-rotbands} show that this is a good
approximation for the energy eigenvalues of the full Heisenberg
model. The simple rotational band approximation predicts a
proportionality constant of $a=0.1$. The linear fits as shown in the
insets give the results $a=0.111$ for $s=1$ and $a=0.108$ for
$s=3/2$, very close to the simple rotational
band approximation. However, if one uses these (DMRG) data to
calculate the zero-temperature magnetization curve, it becomes
clear that there are some crucial deviations from the ideal
parabolic dependence. If there was an ideal parabolic
dependence, the resulting magnetization curve would consist of
steps with constant
widths. Fig.~\ref{fig:ico-magnetization-curves-1} shows the
resulting zero-temperature magnetization curves as calculated
using the DMRG data. Again, the exact diagonalization data for
$s=1/2$ is taken from Ref.~\cite{RLM:PRB2008}. 
 
One can see, that the magnetization curves do not consist of
steps with constant widths. There are some anomalies as expected
for frustrated systems. The plateaus at
$\mathcal{M}/\mathcal{M}_{\rm sat}=1/3$ are clearly visible. The
magnetization jumps due to the independent magnons
\cite{SSRS:EurPhysJB2001} are also visible. Since the jump 
has a height of $\Delta M=3$, it is clear that this effect
vanishes for $s\rightarrow\infty$ in the plot because the
magnetization is normalized by the saturation magnetization
$\mathcal{M}_{\rm sat}=30g\mu_B s$. For $s\rightarrow\infty$,
the classical result, i.e., a strictly linear magnetization
curve \cite{AxL:PRB2001}, will be reached. On the contrary, the plateau
seems to be very stable when increasing $s$. A similar behavior
has already been found for the cuboctahedron, compare
Ref.~\cite{RLM:PRB2008}.  

\begin{figure}[ht!]
	\centering
		\includegraphics[width=0.65\textwidth]{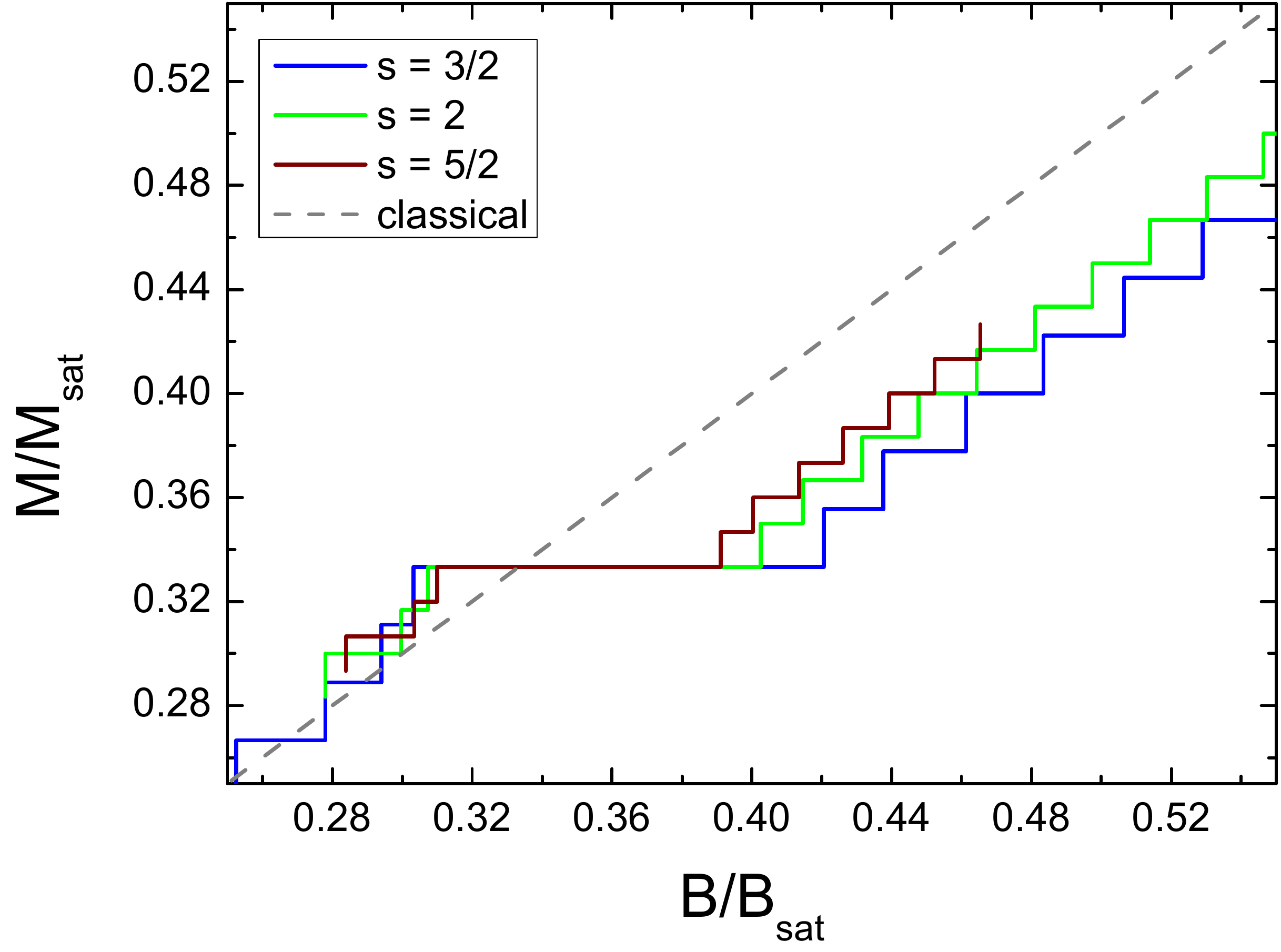}
	\caption{Zero-temperature magnetization curves of the
	icosidodecahedron for $s=3/2,2,5/2$  as obtained by DMRG
	calculations (cf. Fig.~\ref{fig:ico-rotbands}). The plot
	also shows the classical result \cite{AxL:PRB2001}. The
	data is normalized to the saturation field and
	magnetization.} 
	\label{fig:ico-magnetization-curves-2}
\end{figure}

\begin{figure}[ht!]
	\centering
		\includegraphics[width=0.65\textwidth]{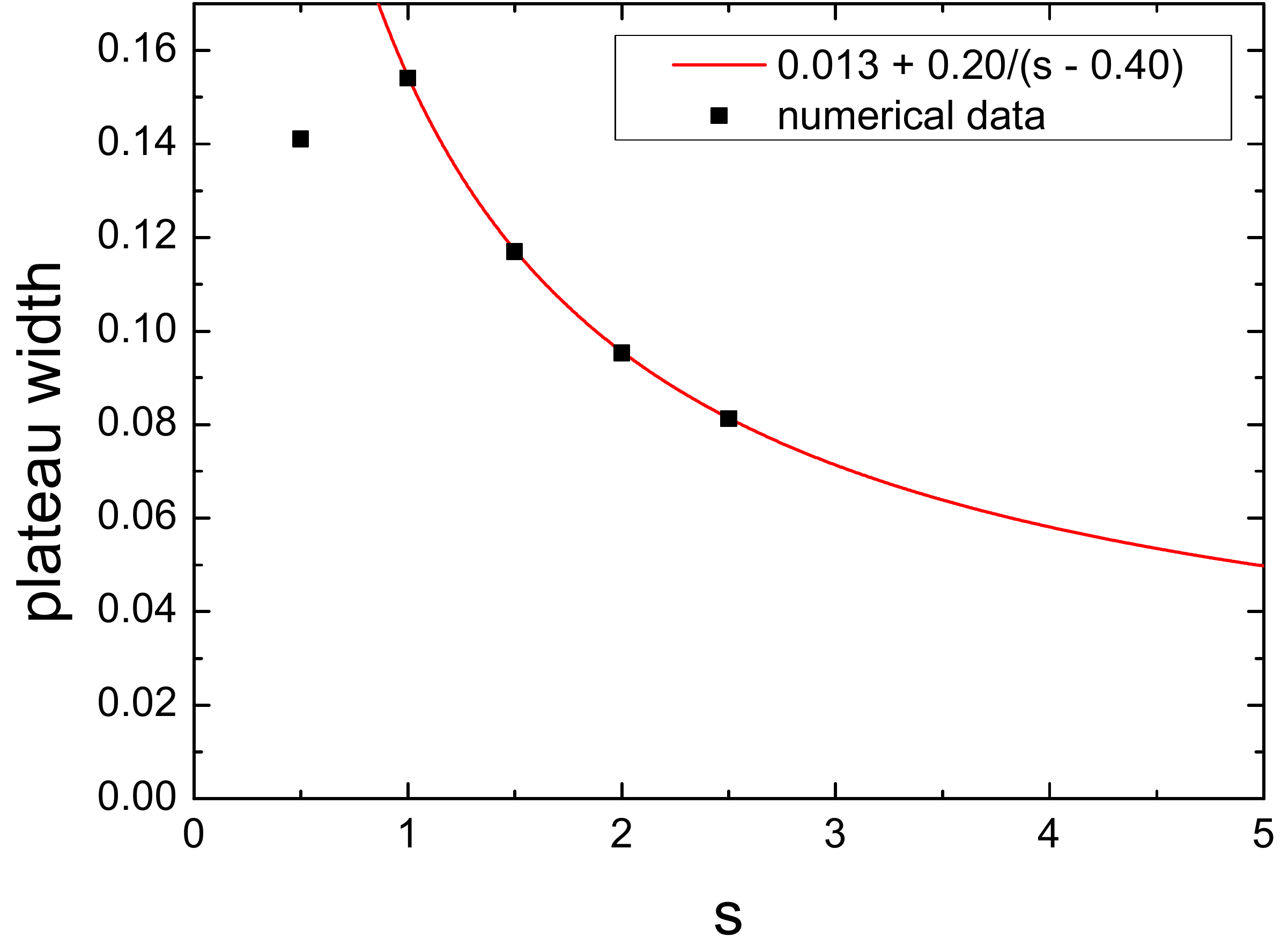}
	\caption{Plateau widths for different $s$
	quantum numbers as obtained by DMRG $S>1/2$ and exact
	diagonalization $s=1/2$ calculations. The exact
	diagonalization values are taken from
	Ref.~\cite{RLM:PRB2008}. This part of the plot also
	includes an extrapolation to $s\rightarrow\infty$ using
	the data for $s>1/2$. The extrapolated value for the
	plateau width is $0.013$. } 
	\label{fig:ico-plateaux}
\end{figure}

For the cases $s=2$ and $s=5/2$ we have calculated the lowest
energy eigenvalues only in some $M$ subspaces, including those
subspaces that are relevant for the calculation of the plateau
width. We have kept $m=2000$ density matrix eigenstates for
these calculations. The plateau is clearly visible, even for
$s=5/2$. However, it is not clear if these plateaus can be
measured in experiments, since for $s=1/2$ the plateau
disappears rather quickly with increasing temperature
\cite{ScW:EurPhysJB2010}. Fig.~\ref{fig:ico-plateaux} shows the
plateau widths for different $s$ as obtained by the DMRG
calculations (cf. Fig.~\ref{fig:ico-magnetization-curves-2}) and
the exact diagonalization for $s=1/2$ \cite{RLM:PRB2008}.  
The left part of Fig.~\ref{fig:ico-plateaux} shows an extrapolation to
$s\rightarrow\infty$ using the data for $s>1/2$. The
extrapolated value for the plateau width is $0.013$ and not
exactly zero but rather close the expected value. It has to be
emphasized here that the plateau widths 
shown in Fig.~\ref{fig:ico-plateaux} are approximate values for
$s>1/2$. The accuracy of the results is analyzed in 
the next subsection.

\subsection{Extrapolation to $m\rightarrow\infty$ and error estimates}

\begin{figure}[ht!]
	\begin{minipage}[c]{0.5\textwidth}
		\includegraphics[width=1.0\textwidth]{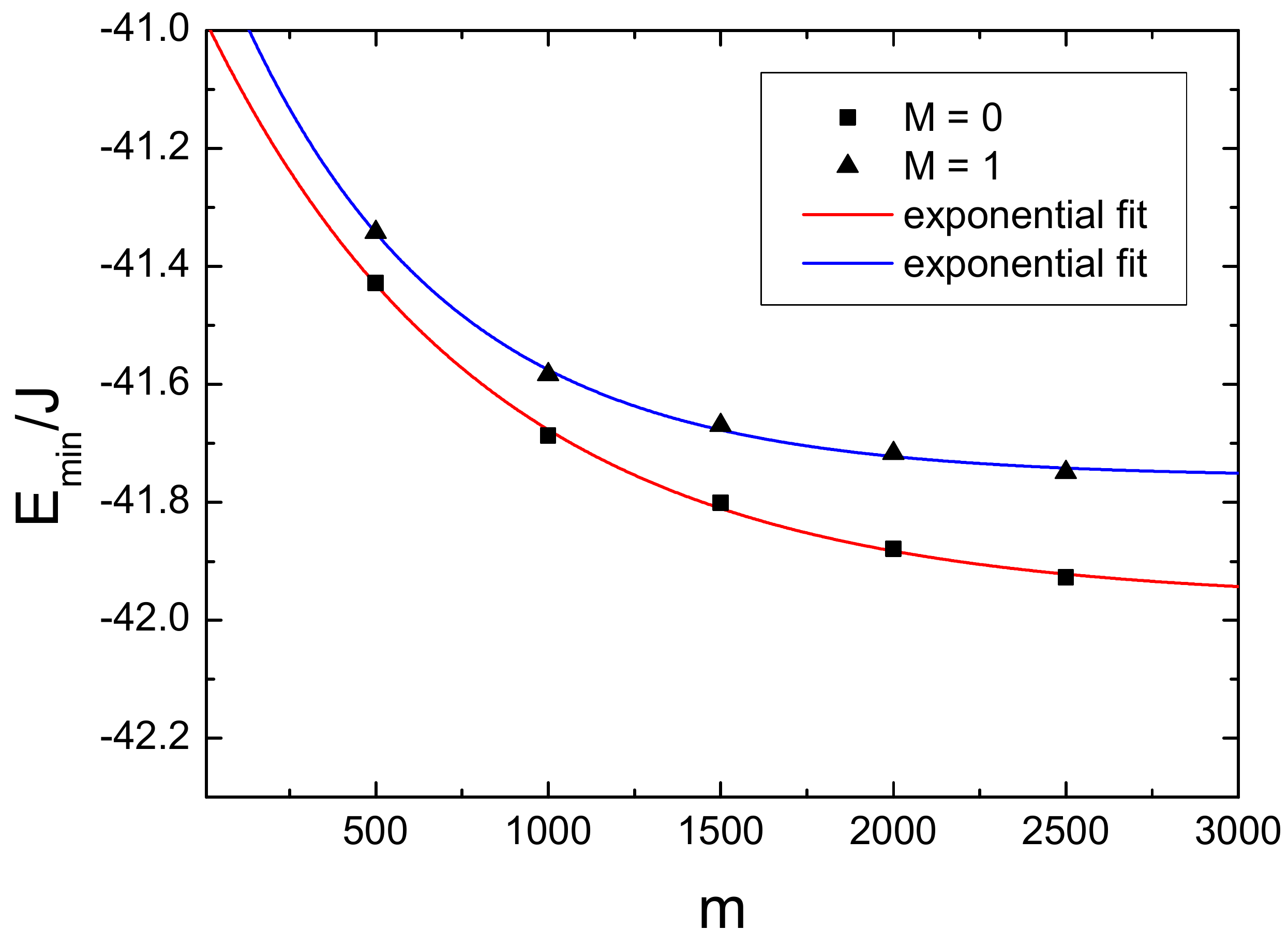}
	\end{minipage}
	\begin{minipage}[c]{0.5\textwidth}
		\includegraphics[width=1.0\textwidth]{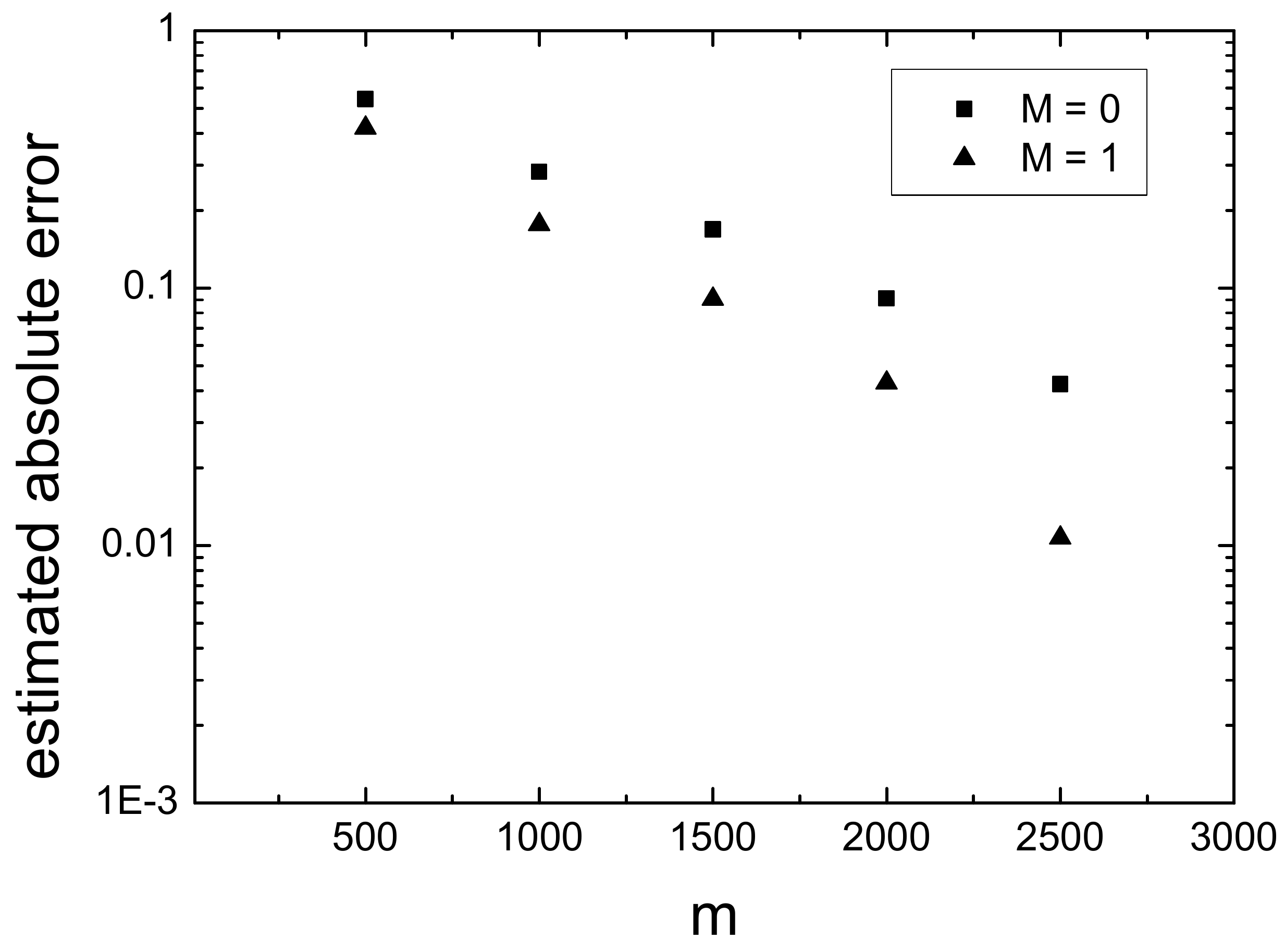}
	\end{minipage}
	\caption{Left: DMRG results for the lowest energy
	eigenvalues of the $M=0$ and the $M=1$ subspace. $m$
	denotes the number of kept density matrix
	eigenstates. The lines are exponential fits to the
	data. Right: estimated absolute error as a function of
	$m$.} 
	\label{fig:ico1-convergence}
\end{figure}

A sensible extrapolation in $m$ (the number of kept density
matrix eigenstates) or $\Delta w$ (truncated weight) is only
possible when the results are fully converged for the current
$m$ value. Fig.~\ref{fig:ico1-convergence} shows the ground
state energy and the lowest energy in the $M=1$ subspace of the
$s=1$ icosidodecahedron as calculated using DMRG for different
$m$ values.

An extrapolation with an exponential ansatz yields $E_{\rm
  min}^{\rm extra}(M=0)=-41.97(2) J$ and $E_{\rm min}^{\rm
  extra}(M=1)=-41.76(1) J$.\footnote{The errors given in
  parenthesis are the standard errors of the fitting procedure.} In
Ref.~\cite{ExS:PRB2003}, the results were extrapolated using an
ansatz of the form $a+b/m$. However, this ansatz did not work
well for our data which is based on much larger $m$ values. The
right part of Fig.~\ref{fig:ico1-convergence} shows the
estimated absolute error, i.e., the difference between the
extrapolated energy and the energy value obtained for finite
$m$, as a function of $m$. For $s>1$ such an extrapolation was
not possible since the production of enough converged
energies for different $m$ values is virtually impossible. 

\begin{figure}[htb!]
    \centering
    \includegraphics[width=0.65\textwidth]{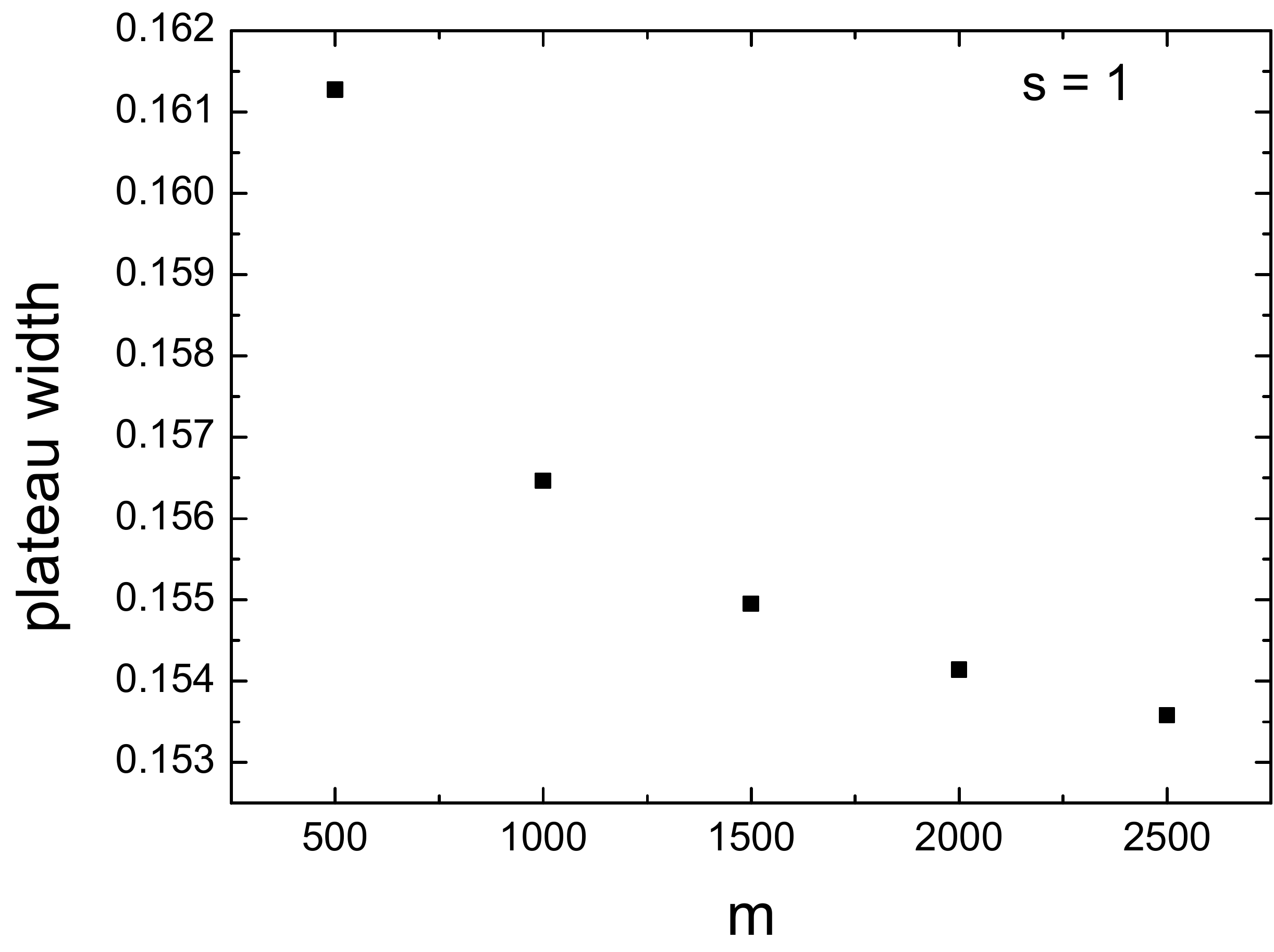}
    \caption{The plateau width of the $s=1$ icosidodecahedron as
    a function of the number of density matrix eigenstates that
    were kept in the DMRG calculations.} 
    \label{fig:ico1-plateau-convergence}
\end{figure}

Fig.~\ref{fig:ico1-plateau-convergence} shows the convergence of
the width of the $\mathcal{M}_{\rm sat}/3$ plateau as a function
of $m$. One can see that the plateau width decreases with
increasing $m$ so that DMRG seems to overestimate the width of
plateau. We find similar effects for $s>1$, but since the
calculations are extremely time-consuming, it was not possible
to obtain enough numerical data for a more systematic study of this
effect. Looking more carefully at the truncated weights, we find
that the DMRG calculations of the lowest energy in the
$M=10s$($=\mathcal{M}_{\rm sat}/(3g\mu_B)$) subspace result in
smaller values of the truncated weight than the calculations in
the adjacent subspaces with the same $m$. However, the order of
magnitude of the truncated weight is still the same. This
indicates that DMRG leads to more accurate results exactly at
one-third of the saturation magnetization which has the
consequence that the method seems to systematically overestimate
the plateau width if one works with the same $m$ value for
adjacent $M$ subspaces. However, we
estimate that the relative errors in the plateau widths are not
larger than approximately 10\%. 

For many one-dimensional systems DMRG can be considered as a
numerically exact method. This is clearly not the case for the
icosidodecahedron. Nevertheless, the width of magnetization
steps would be accurate if the errors of the energy eigenvalues
are approximately the same for adjacent $M$ subspaces. But since
the subspace dimensions become smaller for increasing $M$ it is
clear that calculations with fixed $m$ are 
more accurate for the large-$M$ subspaces. However, we suppose
that the qualitative features of the magnetization curves as
predicted by our DMRG calculations are not altered by these
considerations since the order of magnitude of the errors is
about the same for energy eigenvalues in adjacent
magnetization subspaces.

\subsection{Comparison with CPS and previous DMRG results}

\begin{figure}[ht!]
    \centering
	\includegraphics[width=0.65\textwidth]{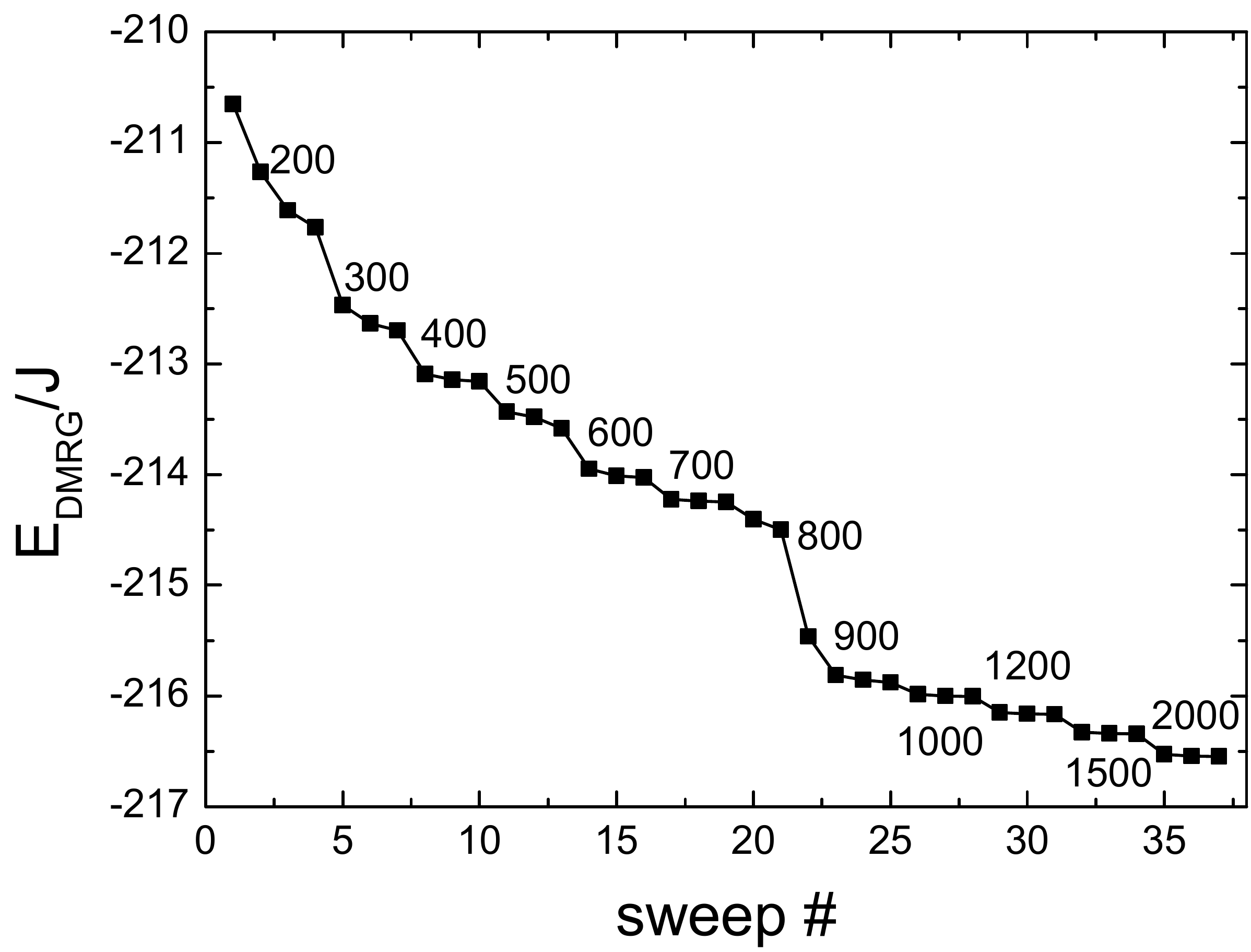}	
	\caption{Ground state energy of the $s=5/2$
	icosidodecahedron as a function of the DMRG sweep. The
	numbers show the retained density matrix eigenstates for
	the current sweep.} 
	\label{fig:ico52-convergence}
\end{figure}

In this subsection we focus on the $s=5/2$ case and the
comparison to previous results on this
system. Fig.~\ref{fig:ico52-convergence} shows the ground state
energy as a function of the DMRG sweep and as a function of the
kept density matrix eigenstates. One can see that even for 2000
kept states and more than 30 sweeps the result is not yet
converged. A much larger number of states is needed to
get convergence. Also, an extrapolation to $m\rightarrow\infty$
is not reliably possible because for that many more sweeps would
have to be performed. 

For $m=2000$ we obtain the value $E_{0}^{\rm DMRG}\approx
-216.5\,J$. This value can be compared with previous
results. The DMRG result of Exler and Schnack for the ground
state energy (with $m=120$) is approximately $-211.1\,J$
\cite{ExS:PRB2003}, a value that is much higher and thus much
more imprecise than our result. The very recent result  of
Neuscamman and Chan using correlator product states in
combination with variational Monte Carlo is $-216.3\,J$
\cite{NeC:arXiv2012}. This value is also higher than our DMRG
result. Also, the comparison of the lowest energies in the $M$
subspaces that are relevant for the calculation of the plateau
width shows that DMRG is -- at least for the CPS ansatz used in
Ref.~\cite{NeC:arXiv2012} -- more accurate if enough states are
used (see Tab.~\ref{tab:dmrg-cps-comparison}).  
\begin{table}[htb!]
	\centering
		\begin{tabular}{c||c|c}
			$M$ & $E_{0}^{\rm DMRG}(M)/J$ & $E_{0}^{\rm CPS}(M)/J$ \\
			\hline
			24 & -158.43 & -154.42 \\
			25 & -153.78 & -149.76 \\
			26 & -147.91 & -144.40 \\
		\end{tabular}
	\caption{Comparison of the DMRG energies to the CPS
	energies. For the DMRG calculations, $m=2000$ states
	were kept. The CPS data is taken from
	Ref.~\cite{NeC:arXiv2012}.} 
	\label{tab:dmrg-cps-comparison}
\end{table}

\subsection{Dynamical correlation function for $s=1/2$}

In the following we investigate the dynamical spin correlations of the 
$s=1/2$ icosidodecahedron in zero external magnetic field. We focus on the
local spin autocorrelation function 

\begin{equation}
S_\mathrm{loc}^{z}(\omega)=-\frac{1}{\pi} \mathrm{Im}
 \langle \mathrm{GS} | S^z_0
\frac{1}{\omega-(H-E_\mathrm{GS}) +i\eta} S^z_0 | \mathrm{GS}\rangle\ .
\end{equation}
This quantity is independent of the specific site, since the ground state is  
found in a non-degenerate, symmetric representation.\footnote{This is different
in some of the ground states realized at finite magnetization, see Ref.~\cite{RLM:PRB2008}.}

\begin{figure}[ht!]
    \centering
\includegraphics[width=0.65\textwidth]{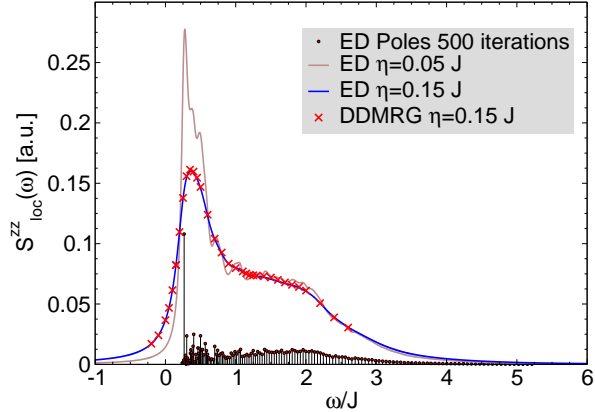}
    \caption{The dynamical correlation function $S^z_\mathrm{loc}(\omega)$
    as obtained by ED and DDMRG calculations. In the ED we have used 500 continued fraction
    iterations whereas in the DDMRG we have used $m=2000$ and
    a broadening $\eta=0.15J$. The agreement between both methods is perfect.
    The vertical lines show the energies where the simple rotational band model predicts the
    peaks.}
    \label{fig:ico12-Szj14}
\end{figure}

We calculate this quantity using two different techniques: i) a 
continued fraction technique based on the Lanczos Exact Diagonalization (ED)
algorithm~\cite{Gagliano1987} and ii) a Dynamical DMRG (DDMRG)
approach \cite{Jec:PRB02}.

In Fig.~\ref{fig:ico12-Szj14} we display the results for $S_\mathrm{loc}^{z}(\omega)$ obtained using both
techniques, with excellent agreement for the same broadening $\eta=0.15J$. The ED results have been
obtained using 500 iterations in the continued fraction expansion, while the DDMRG results have been
obtained using $m=2000$ states, which results in very time-consuming calculations. In the DDMRG $\eta=0.15J$
is fixed during the calculation, while the ED data
can replotted using an arbitrary value for $\eta$. For completeness we have also added the pole content
obtained within the 500 ED iterations. 

Turning to the physics of the spectral function, it displays a rather
sharp peak at approximately $0.35J$ and a very broad shoulder
which then falls off at approximately $2J$. 
The rotational band approximation predicts a two-peak structure 
of the dynamical correlation function. This is clearly not the case since the
``real'' spectrum is much broader and shows only one distinct
peak. However, for such a large $s=1/2$ system, we cannot expect
the rotational band model to be a good approximation. This
approximation is most accurate for small systems with large spin
quantum numbers $s$, compare Ref.~\cite{Wal:PRB2001}.
On the contrary, the strong frustration present in the
icosidodecahedron results in rather dense spectra of low-lying
singlets (even below the lowest triplet), low-lying triplets
(even below the lowest quintet) and so on
\cite{SRS:JMagnMagnMater2005,RLM:PRB2008} of which the
smeared-out dynamical correlation function is a
consequence. This behavior is in accord with the experimental INS
cross sections obtained for the Keplerate \mofe\ with single
spin quantum number $s=5/2$ \cite{Getal:PRB2006}. It would be
very appealing to evaluate the dynamical correlation function
for this compound, unfortunately this is virtually impossible at
the moment. The main problem is that rather accurate lowest
($S=0$) states need to be calculated for the DDMRG method. For the $s=1$
icosidodecahedron, the estimated error of the ground state energy
is already of the order of $0.1J$ using $m=2000$. This inaccuracy would
be too large for a broadening of $\eta=0.15J$ and either a
larger broadening would have to be chosen or more states would
be needed. However, keeping more states strongly increases the
calculation time and a much larger broadening would probably
blur the spectrum too much. Here, a fully optimized and
parallelized DDMRG code would be needed that in addition had to
run on a supercomputer.

\section{Summary}

In this article we report up to date quantum mechanical DMRG
calculations for a class of Keplerate magnetic molecules with
$s=1/2, \dots, 5/2$. We demonstrate that it is possible to obtain
lowest energy 
eigenvalues in orthogonal subspaces and magnetization curves for
$T=0$ with unprecedented accuracy. In addition, using DDMRG as
well as Lanczos techniques, we obtained the dynamical
correlation function for Keplerates with an intrinsic spin of
$s=1/2$.

\section*{Acknowledgment}

The authors thank Eric Jeckelmann, Peter Schmitteckert, Eric
Neuscamman and Garnet Chan for useful discussions. Funding by
the Deutsche Forschungsgemeinschaft (FOR~945)  is thankfully
acknowledged.

\bibliographystyle{elsarticle-num-names}
\bibliography{fe30dmrg.bib}

\end{document}